\documentstyle[eqsecnum,aps,psfig]{revtex}

\def\etal{{\it et\thinspace al.}\ }
\def\eion{{(e~+~ion)}\ }
\def\efe17{{\rm (e~+~Fe~XVII)}}
\def\fe17{{\rm Fe~\sc xvii}}

\def\bprm{{Breit-Pauli R-matrix}\ }
\def\apjl{{Astrophys. J. (Lett.)}}
\def\lr{{$\longrightarrow$}\ }

\def\ll{{$\lambda$}\ }
\newcommand{\abi}{{\it ab~initio}\ }

\newcommand{\rC}{{\rm 3C}}
\newcommand{\rD}{{\rm 3D}}
\newcommand{\rE}{{\rm 3E}}

\newcommand{\ro}{{\rm o}}
\begin{document}
\draft
\preprint{HEP/123-qed}
\title{Influence of Resonances on Spectral Formation of X-Ray
Lines In \fe17}
\author{Guo Xin Chen and Anil K. Pradhan}
\address{
Department of Astronomy, The Ohio State University, Columbus, Ohio
43210\\
}
\date{\today}
\maketitle
\begin{abstract}
  New theoretical results from large-scale relativistic close coupling
calculations reveal the precise effect of resonances in collisional
excitation of {\sc x}-ray lines of Ne-like \fe17. Employing the Breit-Pauli
R-matrix method and a 89-level eigenfunction expansion including up to
$n = 4$ levels shows significant resonance enhancement of
the collision strengths of forbidden and intercombination transitions.
The present results
differ from all previous calculations, heretofore without detailed
resonance structures,
and should help resolve longstanding discrepancies. In particular,
the present line ratios of three benchmark diagnostic lines 3C, 3D, and 3E
at 15.014, 15.265, and 15.456 $\AA$ respectively,
are in excellent agreement with two independent measurements on
Electron-Beam-Ion-Traps [Laming \etal, Astrophys.~J {\bf 545}, L161~(2000)
and Brown \etal, Astrophys.~J {\bf 502}, 1015~(1998)]. The strong
energy dependence due to resonances in these and other cross sections
is demonstrated for the first time. It is of general importance and
strongly manifests itself in {\sc x}-ray plasma diagnostics.

\end{abstract}

Prominent \fe17 {\sc x}-ray lines have long been observed from laboratory and
astrophysical sources [e.g. \cite{rs,smi,hut}],
most recently in the {\it Chandra} and the {\it
XMM-Newton} spectra of active galactic nuclei,
{\sc x}-ray binaries, supernovae, and flaring and non-flaring active stellar
coronae \cite{can,sab}.
 In spite of a number of theoretical, experimental, and observational
studies over decades there remain outstanding
discrepancies related to atomic processes and astrophysical effects
responsible for \fe17 spectral formation in high temperature sources [1-15].
However, the interpretation of experimental and observational ratios of line
intensities usually relies on collisional-radiative (C-R) models using
theoretical cross sections that neglect the fundamental role of
resonant excitation, which preferentially affects the forbidden and
intercombination transitions as opposed to dipole allowed ones.
Of particular interest are three prominent {\sc x}-ray transitions to the
ground level 1s$^2$2s$^2$2p$^6$~$^1$S$_0$ from excited levels:
3C ($\lambda$~15.014$\AA$)
1s$^2$2s$^2$2p$^5$[1/2]3d$_{3/2}$~$^1$P$^\ro_1$
(level 27),
3D $\lambda$~15.265$\AA$
1s$^2$2s$^2$2p$^5$[3/2]3d$_{5/2}$~$^3$D$^\ro_1$ (level
23), and
3E $\lambda$~15.456$\AA$:
1s$^2$2s$^2$2p$^5$[3/2]3d$_{5/2}$~$^3$P$^\ro_1$ (level
17), (level numbers denote energy order relative to the ground level 1).
While the 3C is dipole allowed, the
3D and 3E are spin-forbidden intercombination transitions.
 All but one of the previous calculations used the distorted wave
(DW) approximation that neglects channel coupling, and hence resonances \cite{zha,gol,chem,bha}.
A point of confusion has been that the only previous coupled
channel (CC) calculation \cite{moh}
also yielded the same results as the DW calculations \cite{bha}, because the
CC R-matrix calculations were at energies {\it above all target levels
in the eigenfunction expansion}; and therefore resonances were not
included. Since both the DW and the CC results agreed, it has hitherto
been assumed that resonance effects are not important, and that the cross
sections are relatively constant with energy (e.g. \cite{bro,bro1}).
On the other hand,
the experiments on Electron-Beam-Ion-Traps (EBIT) at Lawrence Livermore
National Laboratory (e.g. \cite{bro}), and at the National Institute of
Standards and Technology (NIST, \cite{lam}),
do not measure absolute excitation cross sections
directly. Rather, they measure relative intensity ratios R1=3C/3D
and R2=3E/3C at a few selected energies. Therefore the presence
of resonances, and more generally the energy variations, are not readily
discernible in experimental data (discussed later). The measured
values differ from theoretical ones by up to 50\% for R1, and a factor of
2 for R2 \cite{lam,bro,bro1,bro2,can}. In order to accout for these discrepancies several
mechanisms based on atomic and astrophysical effects have been
put forward, such as polarization, resonance scattering, and
dielectronic satellite blending, which may be of varying importance
under appropriate conditions \cite{lam,bro,sab,wal}.

 However, as evident from the results in this {\it Letter},
there are extensive resonance
structures in cross sections for all transitions in \fe17 due to many
Rydberg series of resonances converging on to a number of $n=3$ and $n=4$
levels. Many infinite and interacting series of resonances arise due to
coupling between open and closed scattering channels
which, in principle, must be included in order to obtain the cross
sections precisely. But for a complex ion such as \fe17 the number
of channels is very large and the CC calculations become enormously
difficult, especially since relativistic fine structure
must also be considered in addition to other atomic effects. Using the
Breit-Pauli R-matrix (BPRM) method \cite{ip,ben95} we construct a large
eigenfunction
expansion including 89 levels corresponding to 49 LS terms up to the
$n=3$ and the $n=4$ complexes of \fe17. Full details of the
calculations will be presented elsewhere, but we briefly outline these below.
The  coupled-channel wavefunction expansion for the
\efe17 system may be expressed as
$\Psi(E; {\rm e + \fe17}) = \sum_{i} \chi_{i}(\fe17)\theta_{\rm e}(\ell_i) +
\sum_{j} c_{j} \Phi_{j}({\rm Fe}~{\sc xvi})$,
where the $\Psi$ denote the continuum (E $>$ 0)
states of given each total angular momemtum and parity J$\pi$,
expanded in terms of the core ion eigenfunctions
$\chi_i(S_iL_iJ_i)$; the $\theta_{\rm e}(\ell_i)$ refer to the free-electron
partial wave, and the $\Phi_j$ are short-range correlation functions that
also serve to compensate for orthogonality constraints. The 89 levels
belong to the configurations $
2s^22p^6 ,\ 2s^22p^5 (3s,3p,3d),\ 2s^22p^5 (4s,4p,4d,4f),
\ 2s^12p^6 (3s,3p,3d),\ 2s^12p^6 (4s,4p,4d,4f)$. We consider total
symmetries 2J$ \leq 51$ of both parities explicitly in the BPRM
calculations. The size of the calculations (possibly the largest
\eion scattering calculations to date) may be gauged by the fact
that the dimension of the Hamiltonian matrices ranges up to 10286, for J
= 3.5 with 395 free channels and 486
bound channels ($\Phi_j$) (the largest number of free channels per symmetry
is 401); 25 continuum basis functions are used to
represent the $\Psi(\rm e~+~\fe17)$ in the
inner R-matrix region. Relativistic distorted wave (RDW) and
Coulomb-Born-Bethe approximations are employed to `top-up' the partial
wave summations. Particular attention is paid to the resolution of
resonances, with cross sections computed at about 20,000 energies.

 Figs.~1(a-c) display the dense resonance structures in
excitation collision strengths for the 3C, 3D, and 3E transitions
respectively. The results are compared with the RDW values (filled
circles, \cite{zha}) that
essentially represent previous DW calculations (such as from
the HULLAC code \cite{gol,chem}), and the one {\it non-resonant} value from
the previous CC R-matrix calculation (filled square, \cite{moh}). As seen
from the figures, the other
data correspond roughly to the `background' collision strength compared
to the present detailed results. The energy behavior of the three
transitions may be ascertained from the eigenfunction expansions of the
upper levels: 3C: $0.7857|27\rangle+0.1753|23\rangle+0.0305|17\rangle$,
3D: $0.7479|23\rangle+0.2010|27\rangle+0.0491|17\rangle$, and 3E:
$0.9150|17\rangle+0.0767|23\rangle+0.0030|27\rangle$, where the eigenkets
correspond to the energy level indices. As the mixing coefficients
indicate the 3D has a significant component of $|^1$P$^{\rm o}\rangle$, but
the 3E considerably less so. Therefore the 3D would depart from LS coupling
to intermediate coupling to jj-coupling schemes along the neon
isoelectronic sequence with Z, {\it and with $\Omega(\rD)$ increasing with
energy} like $\Omega(\rC)$.
On the other hand the 3E remains largely a (spin) forbidden
transition and $\Omega(\rE)$ decreases with energy. The Einstein A-coefficients
of the 3C, 3D, and 3E are $2.47 \times 10^{13}, 6.01 \times 10^{12}$, and
$9.42 \times 10^{10}$ sec$^{-1}$ respectively. Other interesting atomic
physics aspects of these and other transitions in neon-like ions will be
discussed in later publications, with particular reference to laser
transitions.

Spectral formation in laboratory and astrophysical plasmas needs
to be distinguished from each other. While we describe the latter using a
C-R model for \fe17, generally using
collision strengths averaged over a Maxwellian characterizing a
temperature, the former are measured using a mono-energetic beam with a
certain veocity spread assumed to be a gaussian. We compute the
collision strengths averaged three ways: (i) Maxwellian, (ii) gaussian,
and (iii) numerical. Present results for (ii) and (iii) are shown in
Figs.~1(a-c). In order to compare
with EBIT experiments the gaussian FWHM is
taken to be 30 eV. We note that the energy variations, dependent
on resolution and density of resonances in a given energy region, are
reflected in the gaussian averages (ii); the numerical averages are
slowly varying.
The theoretical line ratios using gaussian and numerical
averages $\langle\sigma\cdot v\rangle$  (GA and NA respectively; $\sigma$
denotes the cross section), are computed taking account of the
radiative branching ratios for 3C, 3D, and 3E which are: 1.0, 1.0, and 0.89
respectively. Therefore we have
$R1=\rC/\rD=\langle\sigma_{\rC}\cdot v\rangle/\langle\sigma_{\rD}\cdot
v\rangle$ and
$R2=\rE/\rC= 0.89 \langle\sigma_{\rE}\cdot v\rangle/\langle\sigma_{\rC}\cdot
v\rangle$.
The results are compared with several EBIT measurements in Table I.
These direct
theoretical results agree with measured values to within experimental errors.
Although our calculations extend up to the $n = 4$ levels, the
highest threshold (level 89) is at $\sim$84.5 Ryd or 1.15 keV;
no resonances above
this energy are therefore included. While the line ratios at 0.85 keV and
0.9 keV are computed directly from the collision strengths as in Figs.~1(a-c),
some extrapolation of the averaged values is made to compare with
the experimental value at 1.15 keV since we expect similar resonance
enhancement above the $n = 4$ due to still higher thresholds.
The GA results differ rather more
than the NA results but still agree with experiment, in contrast to
all other theoretical values. However,
there is significant variation with energy in the GA collision strengths
as they `oscillate' irregularly with energy depending on the density of
resonances prevalent in a given range. This
appears to be reflected in similar `oscillatory' structure seen in the
experimental values reported in \cite{bro2} throughout the energy range
0.1 - 4 keV.
\begin{table}
\caption{Comparison of present line ratios for R1=3C/3D and R2=3E/3C with EBIT
measurements}
\begin{center}
\begin{tabular} {ccccc}
&&$E_i$=0.85 keV&0.9 keV& 1.15 keV\\
\hline
&EBIT&2.77$\pm$0.19$^a$&2.94$\pm$0.18$^b$&(3.15$\pm$0.17,2.93$\pm$0.16)$^a$\\
R1=3C/3D&Theory$^c$:NA&2.80&3.16&3.05$^\dag$\\
&Theory$^c$:GA&2.95&3.27&3.10$^\dag$\\
&Other Theory&\multicolumn{3}{c}{3.78$^d$;4.28$^e$;3.99$^f$}\\
\hline
&EBIT&&0.10$\pm$0.01$^b$&\\
R2=3E/3C&Theory$^c$:NA&0.11&0.085&0.07$^\dag$\\
&Theory$^c$:GA&0.11&0.083&0.07$^\dag$\\
&Other Theory&\multicolumn{3}{c}{0.04$^d$;0.05$^e$;0.05$^f$}\\
\end{tabular}
\end{center}
$^a$ EBIT experiments at LLNL \cite{bro};
$^b$ EBIT experiments at NIST \cite{lam};
$^c$ present theory with NA and GA;
$^d$ \cite{zha};
$^e$ \cite{bha};
$^f$ \cite{moh};
$^\dag$present values with extrapolation of resonance enhancement from
\abi collision strengths from E $\leq$ 1.02 keV (see text).
\end{table}
We also find that it is the $n = 4$ resonances, rather than the $n = 3$,
that dominate the enhancement of electron impact cross sections in
\fe17. The $n = 4$
resonances begin with resonant configurations $2s^22p^53\ell4\ell'$
that manifest themselves from $\sim$47 Ryd, considerably {\it
below} the excitation thresholds of the 2p \lr 3d lines 3C, 3D, and 3E
at $\sim$60 Ryd. Therefore there are relatively fewer $n = 3$
resonances and the $n = 4$ resonances greatly influence the
near-threshold behavior of these cross sections. In the region
75 - 84.5 Ryd the resonances are
due to thresholds corresponding to two-electron-excitation
configurations $2s2p^64\ell$ with weakly coupled channels and
resonances.
This region is expected to be dominated by $n > 4$ resonances.
It might also be mentioned that radiation
damping of autoionizing resonances in cross sections
of \fe17 is known to be negligible \cite{pz97,petal,znp}.

 We note a few specific cases that exemplify the
findings in this {\it Letter}. An application of the present rates would
be to estimate more preciesly the
degree of resonance scattering of the 15.014 $\AA$ line in the solar
and stellar coronae as discussed in \cite{lam,bro1}. Based on our
C-R model with the present collisional data and
extensive new radiative calculations for \fe17 transition
probabilities using the BPRM method (as in \cite{np})
and from the SUPERSTRUCTURE code \cite{eis}, we obtain the 3C/3D ratio
to vary between 2.63 - 3.20, which agrees with the values from EBIT
and from flaring solar corona, but is still higher than the
observed value \cite{wal} of 1.87$\pm$0.21 from non-flaring active
region. This further suggests that
resonance scattering of 3C, or possibly some other mechanism, may be operative
under astrophysical conditions \cite{lam}. Another example is the
discussion of the diagnostic
utility of the 3C/3D line ratio by Brown \etal \cite{bro1} who (a) assume the
cross sections to be relatively constant, and (b) mutiply the RDW cross
section of Zhang and Sampson by 1.25 to agree with the measured value.
From the present work neither (a) nor (b) are necessary. Furthermore,
Brown \etal obtain the temperature dependence of 3C/3D including an
inner-shell satellite line blended with the 3D, but using constant
cross sections. A revised analysis with the present
cross sections should yield a different temperature diagnostics.

 More generally, the excitation of most of the
$n = 3$ levels of \fe17 is similarly affected. Using our
89-level C-R model, with newly computed collsional and radiative
data employing the BPRM method, we investigated the prominent transitions
corresponding to the `coronal' {\sc x}-ray lines $2s^22p^53s$ \lr $2s^22p^6$ at
\ll\ll $16.780, 17.055$ and $17.100~\AA$ (e.g. \cite{mlf}).  Fig.~2(a) shows the
detailed collision strength for the forbidden J = 1 \lr 0 \ll 16.780 $\AA$
(3F) transition, compared with previous DW values (filled circles
\cite{zha} and square \cite{bha}).
The forbidden transitions are expected to be most enhanced by resonances,
as in Fig.~2(a), since the background cross sections are much smaller than
for allowed transitions.
The temperature dependence of the forbidden (3F) to the allowed (3C)
line ratio R3 = 3F/3C = I(16.780)/I(15.014)
is demonstrated in Fig.~2(b), and compared
with that calculated using DW cross sections (filled
squares \cite{bha}); at higher temperatures the resonance
contribution decreases progressively with energy due to the Maxwellian
factor. The electron density dependence is small;
solid-line and dot-line correspond to 10$^{13}$ and 10$^9$ cm$^{-3}$
respectively. The 4 open
circles with error bars are observed and experimental values.
At all temperatures T $< 10^7$ K the present line ratio departs considerably
from those using DW data without resonances, to more than a factor of 3
at about 10$^6$ K---a fact of considerable importance in
photoionized {\sc x}-ray plasmas that have
temperatures of maximum abundance much lower than that in coronal
equilibrium T$_m \sim  4 \times 10^6$ K for \fe17, as marked.
\begin{table}
\caption{Rate coefficients at
$T_{\rm e}$=200 eV for resonant excitation to the $2s^22p^53s$ levels}
\begin{center}
\begin{tabular} {ccccc}
final&\multicolumn{4}{c}{Rate Coefficient ($\times10^{-13}$cm$^3$sec$^{-1}$)}\\
state&Present&Smith \etal~\cite{smi}&Goldstein \etal~\cite{gol}&Chen\&Reed~\cite{chem}\\
\hline
$2p^53s~^3$P$^{\rm o}_1$&22.7&48.0&15.12&14.2\\
All 2p$^5$3s states&70.4&147.0&46.68&42.7
\end{tabular}
\end{center}
\end{table}
Resonant excitation may be considered indirectly by approximate methods
such as quantum
defect theory \cite{znp} or isolated resonance approximations (IRA)
\cite{smi,gol,chem}.
Table II compares the present rate coefficients for the individual 3F line,
and for all transitons to the $2s^22p^53s$ levels, with IRA calculations
using HULLAC \cite{gol}, multi-configuration Dirac-Fock \cite{chem}, and
semi-relativistic Hartree-Fock \cite{smi} methods.
The present data are
$\sim$50\% higher and differ from \cite{gol,chem} which
(i) neglect interference between resonances, and (ii)
take limited account of the several infinite Rydberg series of
resonances resulting in a considerable underestimate of resoance enhancement.
The imprecision of the IRA is also indicated by more than a factor of 3
discrepancy between \cite{chem} and \cite{smi}, both
using IRA but with different decay channels. Present results
are about a factor of 2 lower than \cite{smi}.
Although resonances are generally less important for highly charged ions,
the resulting enhancement in the BP R-matrix cross sections for 
forbidden and intercombination lines, compared to the allowed line, 
yields the ratios R1 and R2 closer to experiments compared to other works.
We also note that ionization and recombination processes are not
included in the present C-R model; dielectronic recombination from
Fe~XVIII to Fe~XVII may also contribute to line emissions.

 The main conclusions of this {\it Letter} are:
(I) resonances are present in \fe17 collision cross sections at all
energies, (II)
the energy dependence of different types of forbidden, intercombination,
and allowed spectral transitions is demonstrated, (III) the effect on
diagnostic line ratios is significant, and may be up to several factors
compared to previous data neglecting resonance effects, (IV)
 the present work is expected to be generally applicable to the
temeprature and density analysis
of many astrophysical and laboratory {\sc x}-ray sources (especially
photoionized plasmas and forbidden lines) not only for \fe17
but also other neon-sequence ions of importance in {\sc x}-ray
lasers \cite{ros}, (V) this work is part of a large-scale
program to compute excitation, photoionization, \eion recombination
\cite{petal,znp}, and
transition probabilities of \fe17 using the accurate BPRM method; for
example, a set of A-values for $\sim$30~000 transitions
have been computed.

We would like to thank Dr.~Werner Eissner for assistance with
the Iron Project BPRM codes.
This work was partially supported by the National Science Foundation
and the NASA Astrophysical Theory Program. The computational work was
carried out utilising well over 1000 CPU hours on the Cray YMP, T94, and
SV1 at the Ohio Supercomputer Center in Columbus Ohio.

\def\amp{{Adv. At. Molec. Phys.}\ }
\def\apj{{ Astrophys. J.}\ }
\def\apjs{{Astrophys. J. Suppl.}\ }
\def\apjl{{Astrophys. J. (Lett.)}\ }
\def\aj{{Astron. J.}\ }
\def\aa{{Astron. Astrophys.}\ }
\def\aas{{Astron. Astrophys. Suppl.}\ }
\def\adndt{{At. Data Nucl. Data Tables}\ }
\def\cpc{{Comput. Phys. Commun.}\ }
\def\jqsrt{{J. Quant. Spectrosc. Radiat. Transf.}\ }
\def\jpb{{J. Phys. B}\ }
\def\pasp{{Pub. Astron. Soc. Pacific}\ }
\def\mn{{Mon. Not. R. Astron. Soc.}\ }
\def\pra{{Phys. Rev. A}\ }
\def\ps{{Phys. Scr.}\ }
\def\prl{{Phys. Rev. Lett.}\ }
\def\zpds{{Z. Phys. D Suppl.}\ }
\def\adndt{At. Data Nucl. Data Tables}

\newpage
\begin{figure}
\centering
\psfig{figure=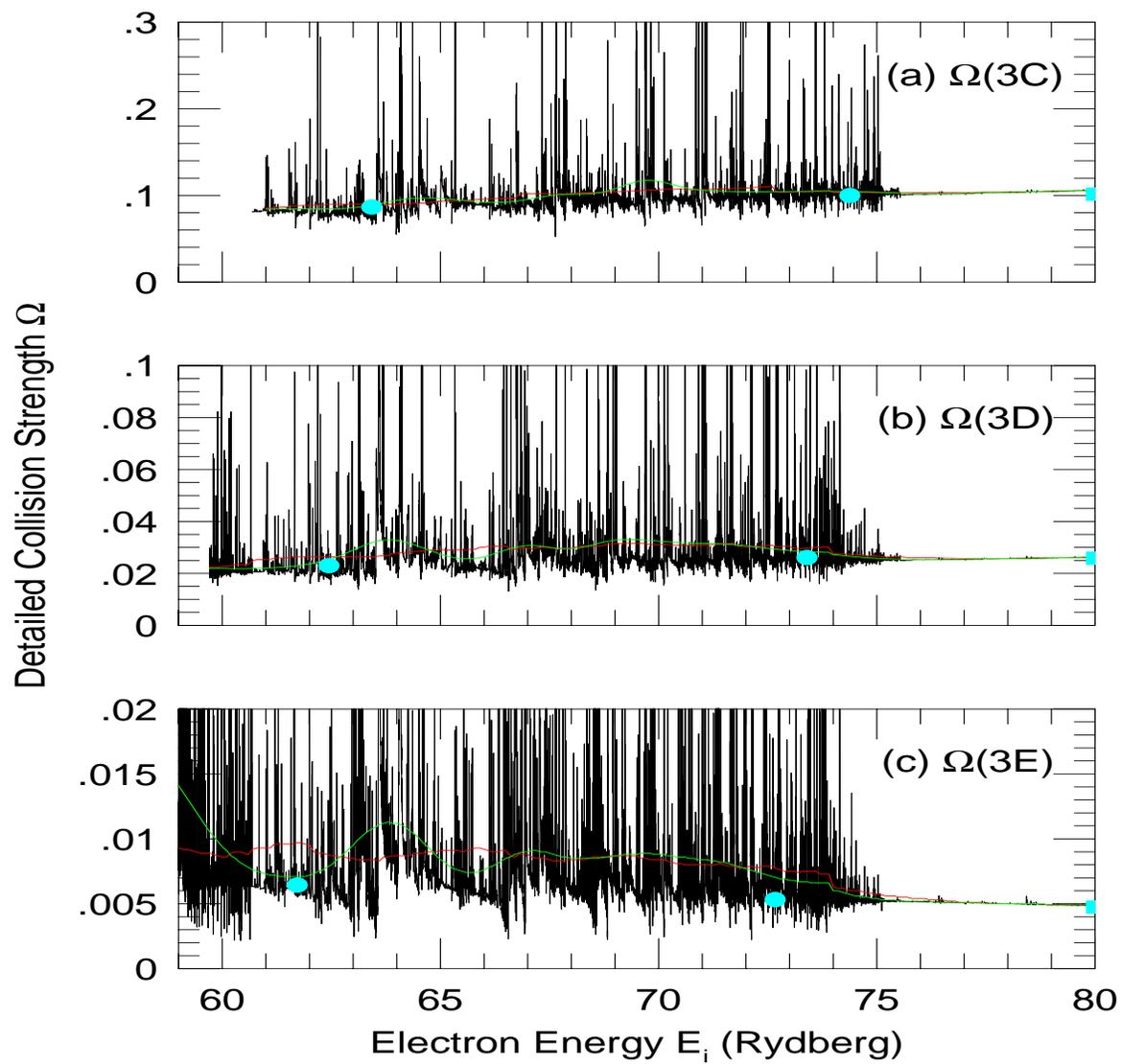,height=17.0cm,width=18.0cm}
\caption{\bprm collision strength $\Omega$ for 3C, 3D and 3E lines
with detailed resonance structures as a function of incident electron energy;
filled dots are RDW results. The gaussian (FWHM = 30 eV) and
numerical averages are also shown (see text).}
\end{figure}

\begin{figure}
\centering
\psfig{figure=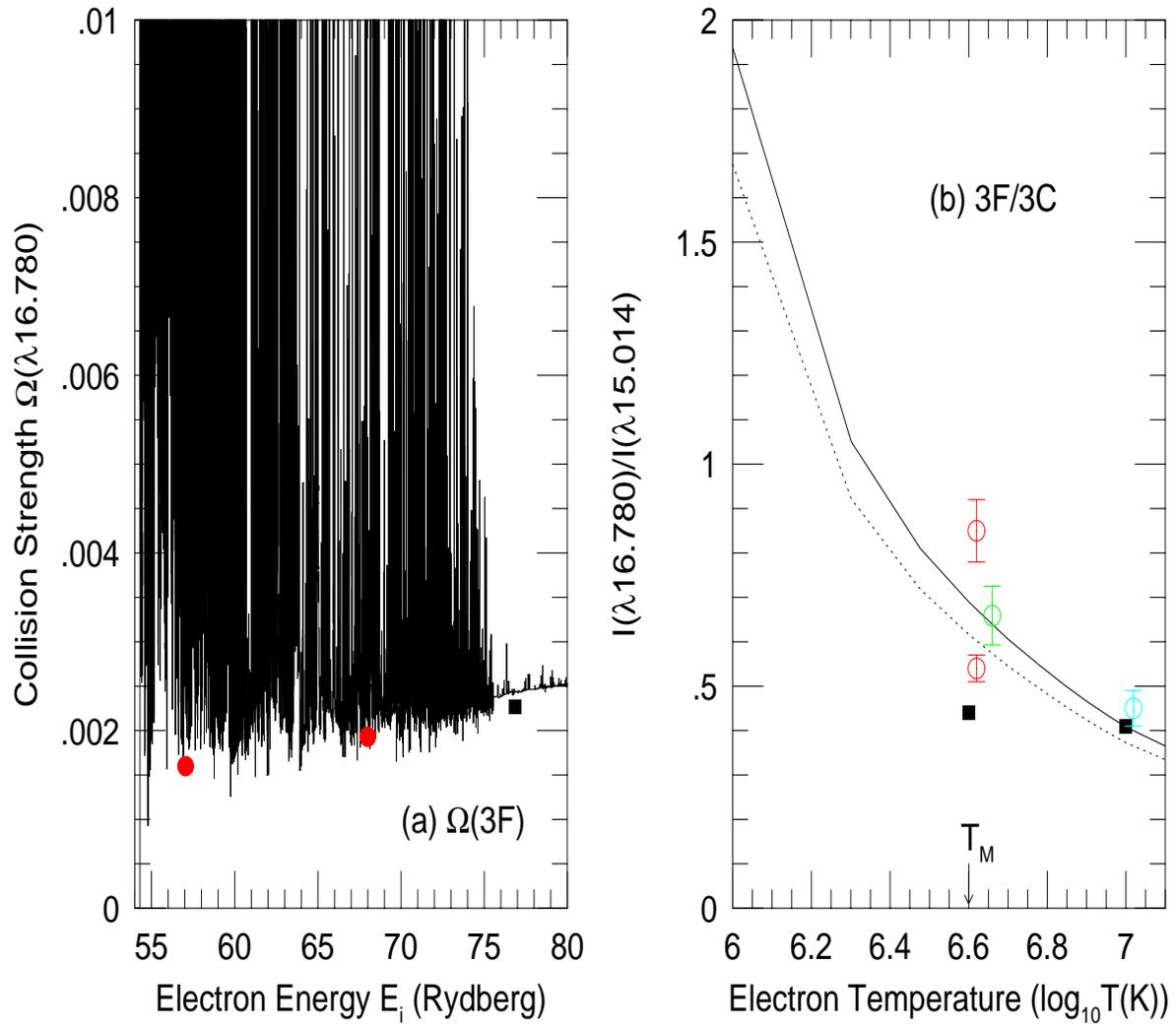,height=17.0cm,width=18.0cm}
\caption{(a): \bprm collision strength $\Omega$ for the 3F line;
the filled circles and square are non-resonant DW calculations;
(b): line ratio 3F/3C vs.~T from a 89-level C-R model.
The electron densities for
solid-line and dot-line curves are 10$^{13}$ and 10$^9$ cm$^{-3}$ respectively.
The 4 open circles with error bars are observed and experimental values: from the solar corona
at T$_m \sim$~4MK \protect\cite{hut},
from the corona of solar-type star Capella at $\sim$~5MK \protect\cite{can},
and from the EBIT experiment at 0.9~keV (log T = 7) \protect\cite{lam}.
The filled squares are values using DW results \protect\cite{bha}.}
\end{figure}


\end{document}